\documentclass[12pt]{article}

\usepackage{amsfonts,amssymb,amsmath,amscd}
\usepackage[ps,dvips,matrix,arrow,frame,import,curve,color]{xy}
\newcommand{\ra}{\rightarrow}
\newcommand{\bul}{\bullet}
\newcommand{\Tr}{{\rm Tr}}
\newcommand{\tep}{{\tilde \epsilon}}

\newcommand{\CC}{{\mathbb C}}
\renewcommand{\th}{{\theta}}
\newcommand{\sm}{{\sigma}}
\newcommand{\Ga}{{\Gamma}}
\newcommand{\Om}{{\Omega}}
\renewcommand{\d}{{\rm d}}
\newcommand{\bi}{{\bar i}}

\newcommand{\tQ}{{\tilde{Q}}}
\newcommand{\J}{{\mathcal J}}
\newcommand{\I}{{\mathcal I}}
\newcommand{\G}{{\mathcal G}}
\newcommand{\cH}{{\mathcal H}}
\newcommand{\cL}{{\mathcal L}}
\newcommand{\cO}{{\mathcal O}}
\newcommand{\al}{{\alpha}}

\newcommand{\la}{{\lambda}}
\newcommand{\no}{\nonumber}
\renewcommand{\part}{\partial}
\newcommand{\TM}{{T\!M}}
\newcommand{\Qb}{Q_{\rm BRST}}
\newcommand{\om}{\omega}
\renewcommand{\i}{\iota}
\newcommand{\op}{\oplus}
\newcommand{\ii}{{i}}
\newcommand{\bpart}{{\bar\partial}}
\newcommand{\bpartial}{{\bar\partial}}
\newcommand{\tI}{{\tilde I}}

\title{Topological sigma-models with $H$-flux and twisted generalized 
complex manifolds}

\author{Anton Kapustin and Yi Li\\{\small \it California Institute of 
Technology, Pasadena, CA 91125, U.S.A.}}

\begin{document}

\begin{titlepage}

\maketitle

\begin{abstract}

We study the topological sector of $N=2$ sigma-models with $H$-flux. It has been known for a long
time that the target-space geometry of these theories is not K\"ahler and can be described in terms of 
a pair of complex structures, which do not commute, in general, and are parallel with respect to two different connections with torsion. Recently an alternative description of this geometry was found,
which involves a pair of commuting twisted generalized complex structures on the target space.
In this paper we define and study the analogues of A and B-models for $N=2$ sigma-models with $H$-flux
and show that the results are naturally expressed in the language of twisted generalized complex geometry.  For example, the space of topological observables is given by the 
cohomology of a Lie algebroid associated to one of the two twisted
generalized complex structures. We determine
the topological scalar product, which endows the algebra of observables 
with the structure of a Frobenius algebra. We also discuss mirror symmetry for 
twisted generalized Calabi-Yau manifolds.

\end{abstract}

\vspace{-6.5in}

\parbox{\linewidth}
{\small\hfill \shortstack{CALT-68-2514}}
\vspace{6.5in}

\end{titlepage}

\section{Introduction}

It was pointed out by E.~Witten in 1988 that one can construct interesting examples of topological
field theories by ``twisting'' supersymmetric field theories \cite{WittenTop}. This observation 
turned out to be very important for quantum field theory and string theory, since observables in 
topologically twisted theories are effectively computable on one hand and can be interpreted in terms 
of the untwisted theory on the other. In other words, supersymmetric field theories tend to have large 
integrable sectors.

{}From the string theory viewpoint, the most important class of supersymmetric field theories admitting
a topological twist are sigma-models with $(2,2)$ supersymmetry. Usually one considers the case when
the B-field is a closed 2-form, in which case $(2,2)$ supersymmetry requires the target $M$ to be a
K\"ahler manifold. In this case the theory admits two different twists, which give rise to two
different topological field theories, known as the A and B-models.\footnote{More precisely, the B-model makes
sense on the quantum level if and only if $M$ is a Calabi-Yau manifold. For the A-model, the
Calabi-Yau condition is unnecessary.} In any topological field theory,
observables form a supercommutative ring. For the A-model, this ring turns out to be a deformation of
the complex de Rham cohomology ring of $M$, which is known as the quantum cohomology ring. It depends on the
symplectic (K\"ahler) form on $M$, but not on its complex structure. For the B-model, the ring of
observables turns out to be isomorphic to
$$
\op_{p,q} H^p(\Lambda^q T^{1,0} M),
$$
which obviously depends only on the complex structure of $M$. Furthermore, it turns out that all
correlators in the A-model are symplectic invariants of $M$, while all correlators in the B-model
are invariants of the complex structure on $M$~\cite{WittenMirr}.

In this paper we analyze more general topological sigma-models for which $H=dB$ is not necessarily
zero. It is well-known that for $H\neq 0$ $(2,2)$ supersymmetry requires the target manifold $M$
to be ``K\"ahler with torsion'' \cite{GHR}. This means that we have two different complex structures $I_\pm$
for right-movers and left-movers, such that the Riemannian metric $g$ is Hermitian with respect to 
either one of them, and
$I_+$ and $I_-$ are parallel with respect to two different connections with torsion. 
The torsion is proportional to $\pm H$.
The presence of torsion implies that the geometry is not K\"ahler (the forms $\omega_\pm=gI_\pm$ are
not closed). Upon topological twisting, one obtains a topological field theory, and one would like
to describe its correlators in terms of geometric data on $M$. As in the K\"ahler case, there are
two different twists (A and B), and by analogy one expects that the correlators of either model
depend only on ``half'' of the available geometric data. Furthermore, it is plausible that
there exist pairs of $(2,2)$ sigma-models with H-flux for which the A-model are B-model are ``exchanged.''
This would provide an interesting generalization of Mirror Symmetry to non-K\"ahler manifolds.

The main result of the paper is that topological observables can be described in terms of a (twisted)
generalized complex structure on $M$. This notion was introduced by N.~Hitchin \cite{Hitchin} and studied 
in detail by M. Gualtieri~\cite{Gua}; we review it below. One can show that the geometric data $H,g,I_+,I_-$ 
can be repackaged as a pair of commuting twisted generalized complex structures on $M$~\cite{Gua}.
We show in this paper
that on the classical level the ring of topological observables and the topological metric on this
ring depend only on one of the two twisted generalized complex structures. 
This strongly suggests that all the correlators of either
A or B-models (encoded by an appropriate Frobenius manifold) are invariants of only one twisted generalized 
complex structure. Therefore, if $M$ and $M'$ are related by Mirror Symmetry (i.e. if the A-model
of $M$ is isomorphic to the B-model of $M'$ and vice versa), then the appropriate moduli spaces 
of twisted generalized complex structures on $M$ and $M'$ will be isomorphic.

To state our results more precisely, we need to recall the definition of the (twisted) generalized
complex structure (TGC-structure for short). Let $M$ be a smooth even-dimensional manifold, and let $H$
be a closed 3-form on $M$. The bundle $\TM\op \TM^*$ has an interesting binary operation, called the twisted
Dorfman bracket. It is defined, for arbitrary $X,Y\in\Gamma(\TM)$ and $\xi,\eta\in\Gamma(\TM^*)$, as
$$
(X\op\xi)\circ (Y\op\eta)=[X,Y]\op\left(\cL_X\eta-i_Y d\xi+\i_Y \i_X H\right).
$$
It is not skew-symmetric, but satisfies a kind of Jacobi identity. Its skew-symmetrization is called
the twisted Courant bracket and does not satisfy the Jacobi identity. The bundle $\TM\op \TM^*$ also
has an obvious pseudo-Euclidean metric of signature $(n,n)$, which we call $q$. For a detailed discussion
of the Dorfman and Courant brackets and their geometric meaning, see Ref.~\cite{Royt}.

A TGC-structure on $M$ is a bundle map $\J$ from $\TM\op \TM^*$ to itself which satisfies the following
three requirements:
\begin{itemize}
\item $\J^2=-1$.
\item  $\J$ preserves $q$, i.e. $q(\J u,\J v)=q(u,v)$ for any $u,v\in \TM\op \TM^*$.
\item The eigenbundle of $\J$ with eigenvalue $i$ is closed with respect to the twisted Dorfman
bracket. (One may replace the Dorfman bracket with the Courant bracket without any harm).
\end{itemize}

In the special case $H=0$, the adjective ``twisted'' is dropped everywhere, and one gets the notion
of a generalized complex structure (GC-structure). 

To any TGC-structure on $M$ one can canonically associate a complex Lie algebroid $E$ (which is, roughly,
a complex vector bundle with a Lie bracket which has properties similar to that of a complexification
of the tangent bundle of $M$). {}From any complex Lie algebroid $E$ one can construct a differential complex
whose underlying vector space is the space of sections of $\Lambda^\bul(E^*)$
(which generalizes the complexified de Rham complex of $M$). We will show that the space
of topological observables is isomorphic to the cohomology of this differential complex. We will also
write down a formula for the metric on the cohomology, which makes it into a supercommutative
Frobenius algebra. Both the ring structure and the topological metric depend only on one of the
two TGC-structures available.

Even if $H=0$, one can consider the situation when $I_+\neq I_-$. This is possible, for example, when
$M$ is hyperk\"ahler, and there is a family of complex structure parametrized by $S^2$ and compatible
with a fixed Riemannian metric $g$. This case was considered in Ref.~\cite{Kap}, where
the relation with GC-structures was first noted. In this paper we extend the results of Ref.~\cite{Kap}
to the case $H\neq 0$. The relation of twisted generalized complex geometry with $N=2$ supersymmetric
sigma-models has also been studied in Refs.~\cite{Lind1,Lind2}. This subject may also be relevant to
flux compactifications of superstring theories, see e.g. Ref.~\cite{GMPT}.

The organization of the paper is as follows. 
In section 2, we give a brief review of $(2,2)$ supersymmetric sigma models. Our emphasis is on their
relation to generalized complex geometry. In section 3, we construct the topological theories by ``twisting'' $(2,2)$ supersymmetric sigma models with $H$-flux. In particular, we discuss the relevance of the twisted generalized Calabi-Yau condition for our construction. The Ramond-Ramond ground states of the $(2,2)$ theory are studied in section 4. The results of this section are used to prove that the space of topological observables is given by an associated Lie algebroid cohomology. In section 5 we discuss the topological
metric on the space of observables in the topologically twisted theory and write down a formula for tree-level topological correlators, neglecting quantum corrections. In section 6 we discuss possible
quantum corrections due to worldsheet instantons. In section 7 we discuss the implications of our results,
including a possible generalization of Mirror Symmetry to non-K\"ahler manifolds with $H$-flux.

\section{The Geometry of $(2,2)$ Supersymmetric Sigma Models}
\label{sec:N=2}

We start by reviewing $(2,2)$ supersymmetric sigma models and setting up the notation. 
It is well-known that the bosonic sigma model in $1+1$ dimensions with any 
Riemannian target manifold $M$ admits a $(1,1)$ supersymmetric 
extension. The action of the $(1,1)$-extended theory has a superfield formulation:
\begin{equation}
\label{eq:action}
S \;=\; \frac12 \int\d^2\sm\d^2\th 
\big(g_{ab}(\Phi)+B_{ab}(\Phi)\big)D_+\Phi^a D_-\Phi^b.
\end{equation}
Here $\Phi^a = \phi^a + \th^+\psi_+^a+\th^-\psi_-^a + \th^-\th^+ F^a$ are 
$(1,1)$ superfields (components of a supermap from a $(1,1)$ superworldsheet to the bosonic target $M$), 
$g$ is a Riemannian 
metric on $M$, and $H=dB$ is a real closed 3-form on $M$. 
Note that if $H$ defines a non-trivial class 
in $H^3(M)$, then $B$ is only locally well-defined. The 
super covariant derivatives $D_\pm$ are defined by
$$D_+ = \frac{\part}{\part\th^+}+i\th^+\part_+, \qquad D_- = 
\frac{\part}{\part\th^-}+i\th^-\part_-, \qquad 
\part_\pm\equiv\part_0\pm\part_1.$$
The action is invariant under the standard supersymmetry generated by
$$Q_+ = \frac{\part}{\part\th^+}-i\th^+\part_+, \qquad Q_- = 
\frac{\part}{\part\th^-}-i\th^-\part_-.$$

When the target manifold $M$ possesses additional structure, the theory may possess a
larger supersymmetry. For example, it is well-known that when $(M,g)$ is 
K\"{a}hler and $H=0$, the theory has $(2,2)$ supersymmetry. A natural question is 
whether $(2,2)$ supersymmetry implies K\"ahler geometry. This has been answered 
in the negative by S.J.~Gates et al.~\cite{GHR}. It is shown there that the general form 
of a second supersymmetry (as opposed to the standard one generated by 
$Q_\pm$ above) is
$$\tilde\delta \Phi^a = (\tep^+\tilde{Q}_+ +  \tep^-\tilde{Q}_-)\Phi^a = 
\tep^+I_+(\Phi)^a_{\;\;b} D_+\Phi^b + \tep^- I_-(\Phi)^a_{\;\;b} 
D_+\Phi^b,$$
where the tensors $I^a_{\pm b}$ satisfy the following constraints. First, 
the condition that the above transformation generates a separate 
(on-shell) $(1,1)$ supersymmetry, which commutes with the standard one, 
requires $I_+,I_-$ to be
a pair of integrable almost complex structures on $\TM$. Second, the 
invariance of the action
(\ref{eq:action}) requires that the metric $g$ be Hermitian with 
respect to both $I_+$ and $I_-$, and that the tensors $I_\pm$ are 
covariantly constant with respect to certain affine connections with 
torsion.
More explicitly, one must have
$$\nabla ^{(\pm)}_a {I_\pm}^b_{\;\;c} = \part_a{I_\pm}^b_{\;\;c} 
+{\Ga_\pm}^b_{\;\;ad}{I_\pm}^d_{\;\;c}-{\Ga_\pm}^d_{\;\;ac}{I_\pm}^b_{\;\;d}=0,$$
with the affine connections defined by
$${\Ga_{\pm}}^a_{\;bc} = \Ga^a_{\;bc}\pm \frac12 g^{ad}H_{dbc}.$$
Here $\Ga^a_{\;bc}$ is the Levi-Civita connection:
$$\Ga^a_{\;bc} = \frac12 g^{ad}\left(\part_b g_{dc}+\part_c g_{bd}-\part_d 
g_{bc}\right).$$
Note that in general $(g,I_+)$ and $(g,I_-)$ do not define K\"{a}hler 
structures (i.e. the
2-forms $\omega_\pm=gI_\pm$ are not closed) due to the presence of 
torsion in $\Gamma_\pm$.

An interesting special case is when $[I_+,I_-]=0$. In this 
case, one can simultaneously diagonalize $I_\pm$ and one can decompose 
$\TM = {\rm ker}(I_+-I_-)\oplus{\rm ker}(I_++I_-)$. It was shown in 
\cite{GHR} that ${\rm ker}(I_+-I_-)$ is integrable to $N=2$ chiral 
superfields whilst ${\rm ker}(I_++I_-)$ is integrable to twisted chiral 
superfields. Such a manifold is said to admit a product structure defined 
by $P=I_+I_-$. Locally, it is a product of two K\"{a}hler manifolds, but 
globally it is not K\"{a}hler in
general. Another interesting class of examples is provided by 
hyper-K\"{a}hler manifolds, which admit a family of complex structures 
parametrized by $\vec{x}\in S^2$. One may take $I_+$ and $I_-$
to be any two complex structures parametrized by two points 
$\vec{x}_\pm\in S^2$.
We refer to Refs. \cite{GHR,Rocek,IKR,BSVV,LZ} for more details on these 
and related issues.

Remarkably, the geometric data required by generic $(2,2)$ 
supersymmetry are equivalent to
those which define the so-called twisted generalized K\"ahler structure 
\cite{Gua}. Here we briefly recall the definitions which are needed later. 
Let $M$ be a real manifold of dimension $2n$. As already mentioned in the 
introduction, the bundle $\TM\op \TM^*$ has a natural pseudo-Euclidean 
scalar product of signature $(2n,2n)$ which we will denote $q$.
A twisted generalized complex structure on $M$ is a section $\I$ of ${\rm 
End}(\TM\oplus \TM^*)$ which preserves $q$,
satisfies $\I^2=-1$, and such that its $i$-eigenbundle is 
closed with respect to the twisted Courant bracket.
A twisted generalized K\"ahler structure on $M$ is a pair of commuting 
twisted generalized complex structures, $(\J_1,\J_2)$, such that 
$\G=-q\J_1\J_2$ defines a positive-definite metric on $\TM\oplus \TM^*$.
It is shown in Ref.~\cite{Gua} that specifying a twisted generalized K\"ahler 
structure is equivalent to
specifying $g,H,I_+,$ and $I_-$ satisfying the constraints of $(2,2)$ 
supersymmetry. Explicitly, the two commuting twisted generalized complex 
structures can be taken as
\begin{equation}
\label{eq:J1J2}
\J_1 = \begin{pmatrix}\tilde I&-\al\\ \delta\om&-{\tilde I}^t\end{pmatrix}, \qquad 
\J_2 = \begin{pmatrix}\delta I&-\beta\\ \tilde\om&-\delta{I}^t\end{pmatrix}
\end{equation}
with
\begin{align*}
&\tilde I=(I_++I_-)/2, &\delta I=(I_+-I_-)/2\\
&\beta=(\om_+^{-1}+\om_-^{-1})/2, &\al=(\om_+^{-1}-\om_-^{-1})/2\\
&\tilde\om=(\om_++\om_-)/2,  &\delta\om=(\om_+-\om_-)/2
\end{align*}
It can be shown that both $\al$ and $\beta$ are (possibly degenerate) Poisson 
bivectors \cite{LZ}. 

\section{Construction of Topological Theories}

\subsection{Twisting}

In this section, we construct topologically twisted versions of 
$(2,2)$ sigma-models without
assuming $I_+=I_-$. In fact, the case $H=0, I_+\neq I_-$ has already been analyzed in 
\cite{Kap}. It is shown there that the space of local observables of the 
topologically twisted theory can be identified with the cohomology of a certain 
Lie algebroid associated to generalized complex structure. We will analyze
the general case.

We follow the approach pioneered in Ref.~\cite{WittenMirr}, which dealt with 
the case when $M$ is a K\"{a}hler
manifold, with vanishing $H$-field, and $I_+=I_-$. If there is 
a non-anomalous $U(1)$ R-symmetry satisfying
an integrality constraint (for any state the sum of spin and one-half the R-charge
must be integral),
then one may shift the spin of all fields by one-half of their R-charges and 
obtain a topological
field theory. In the case when the target space is a K\"ahler manifold, 
there are two classical $U(1)$ R-symmetries: the vector R-symmetry 
$U(1)_V$ and the axial R-symmetry $U(1)_A$. At the quantum level, $U(1)_V$ 
remains a good symmetry, while $U(1)_A$ suffers from an anomaly unless $M$ 
satisfies the condition $c_1(\TM)=0$. Twisting by the vector 
R-symmetry yields the so-called A-model, while twisting by the axial 
R-symmetry (if it is not anomalous) yields the B-model.

It is not difficult to see that the construction in Ref.~\cite{WittenMirr} can 
be applied to the more general case at hand. The two complex structures, 
$I_\pm,$ induce two different decompositions of the complexified tangent 
bundle
$$\TM_\CC\simeq T_+^{1,0}\oplus T_+^{0,1}\simeq T_-^{1,0}\oplus 
T_-^{0,1}.$$
Under such decompositions the fermionic fields $\psi_\pm$ splits 
accordingly into the holomorphic and anti-holomorphic components:
$$\psi_+=\frac12(1-iI_+)\psi_+ + \frac12(1+iI_+)\psi_+, \quad 
\psi_-=\frac12(1-iI_-)\psi_- + \frac12(1+iI_-)\psi_-.$$
At the classical level, there are two inequivalent ways to assign $U(1)$ 
R-charges to fermions (the bosons
having zero charge):
\begin{eqnarray}
 	U(1)_V: && q_V\Big(\frac12(1-iI_+)\psi_+\Big)=-1, \qquad 
q_V\Big(\frac12(1-iI_-)\psi_-\Big)=-1\no\\
 	U(1)_A: && q_A\Big(\frac12(1-iI_+)\psi_+\Big)=-1, \qquad 
q_A\Big(\frac12(1-iI_-)\psi_-\Big)=1\no
\end{eqnarray}
The topological twisting is achieved by shifting the spin of fermions 
either by $q_V/2$ or $q_A/2$. We will call the corresponding topological 
theories generalized A and B-models. Note that flipping the sign of $I_-$ 
exchanges them.

So far our analysis has been at the classical level. For the generalized A 
and B-models to make sense as quantum field theories, one must require 
that the $U(1)$ symmetry used in the twisting be anomaly-free. The 
anomalies are easily computed by the Atiyah-Singer index theorem, and the 
resulting conditions are
\begin{eqnarray}\label{anomaly}
U(1)_V: && c_1(T_-^{1,0})-c_1(T_+^{1,0}) = 0\\
U(1)_A: && c_1(T_-^{1,0})+c_1(T_+^{1,0}) = 0\no
\end{eqnarray}

It is possible to express the anomaly conditions in terms of twisted 
generalized complex structures.
Recall that $(2,2)$ supersymmetry requires $M$ to be a twisted 
generalized K\"ahler manifold, with two commuting twisted generalized 
complex structures $(\J_1, \J_2)$ and positive definite metric
$\G=-q \J_1\J_2$ on $\TM\oplus \TM^*$. Let $C_\pm$ be the $\pm1$ 
eigenbundles of $\G$. The natural projection from $\TM\oplus \TM^*$ to 
$\TM$ induces bundle isomorphisms $\pi_\pm: C_\pm\simeq \TM$. The twisted
  generalized complex structure $\J_1$ induces two complex structures on 
$\TM$, one from $\pi_+:C_+\to\TM$ and the other from $\pi_-:C_-\to\TM$. 
These are the two complex structures $I_\pm$ which appeared above.
Let $E_1$ and $E_2$ denote the $i$-eigenbundles of $\J_1$ and 
$\J_2$ respectively. Since $\J_1$ and $\J_2$ commute, one has the further 
decompositions $E_1=E_1^+\oplus E_1^-$ and $E_2=E_2^+\oplus E_2^-$, where 
the superscripts $\pm$ label the eigenvalues $\pm i$ of the other 
twisted
generalized complex structure. It follows that
$$C_\pm\otimes\CC = E_1^\pm\oplus(E_1^\pm)^* = E_2^\pm\oplus(E_2^\pm)^*.$$
Now we can rewrite the conditions Eq.~(\ref{anomaly}) in terms of bundles 
$E_1$ and $E_2$:
\begin{eqnarray}\label{anomaly2}
U(1)_V: && c_1(E_2) = 0 \\
U(1)_A: && c_1(E_1) = 0\no
\end{eqnarray}
It seems natural to call either of these conditions a twisted generalized Calabi-Yau condition
(for $\J_1$ and $\J_2$, respectively). However, this name is already reserved for a somewhat stronger
condition introduced by Hitchin~\cite{Hitchin} and Gualtieri~\cite{Gua}. The Hitchin-Gualtieri condition
on $\J$ implies the vanishing of $c_1(E)$, but the converse is not true in general. Physically,
the vanishing of $c_1(E)$ is also not sufficient for the topological twist to make sense.
We will see in sections 4 and 5 that the topological twist makes sense if and only if the Hitchin-Gualtieri
condition is satisfied.

As already mentioned, flipping the relative sign 
of $I_\pm$ exchanges the generalized A- and B-models. This is consistent 
with the anomaly-cancellation condition, since changing the relative sign 
of $I_\pm$ is
equivalent to exchanging $\J_1$ and $\J_2$.

\subsection{BRST cohomology of operators}

Next we describe the BRST operator and BRST-invariant observables. We 
shall focus mainly on the
generalized B-model, since the A-model can be obtained from it by flipping 
the sign of $I_-$. As we just
discussed, $(M,\J_1)$ must be a twisted generalized Calabi-Yau manifold. 
After the topological twist,
$(1+iI_+)\psi_+$ and $(1+iI_-)\psi_-$ become sections of $T_+^{0,1}$ 
and $T_-^{0,1}$ respectively\footnote{Actually these are sections of the {\em pullbacks} of $T_+^{0,1}$ and $T_-^{0,1}$. However we shall slightly abuse the notation by not spelling out explicitly the word ``pullback'' in the following.}. On the other hand, $(1-iI_+)\psi_+$ 
and $(1-iI_-)\psi_-$ become worldsheet spin-1 fields and they should not 
appear in the BRST variation of $\phi$. We obtain two scalar nilpotent 
operators from the original supersymmetry generators: $Q_L = (Q_+ + 
i\tQ_+)/2$ and $Q_R = i(Q_-+i\tQ_-)/2$. The overall factor of $i$ in the expression of $Q_R$ is for later convenience. The $N=2$ supersymmetry algebra 
implies that $Q_L^2=0$, $Q_R^2=0$, and $\{Q_L,Q_R\}=0$. We take the BRST 
operator of the generalized B-model to be $Q_{\rm BRST}=Q_L+ Q_R$.

To simplify notation, let us define:
$$\chi\equiv\frac12(1+iI_+)\psi_+,\qquad  \la\equiv 
\frac{i}2(1+iI_-)\psi_-.$$
The scalar fields transform under $Q_L$ and $Q_R$ as follows:
\begin{eqnarray}
\{Q_L,\phi^a\} &=& \chi^a\no\\
\{Q_L,\chi^a\} &=& 0\no \\
\{Q_L,\la^a\} &=& -{\Ga_-}^a_{\;\;bc}\chi^b\la^c\no \\
\{Q_R,\phi^a\} &=& \la^a\\
\{Q_R,\la^a\} &=& 0\no \\
\{Q_R,\chi^a\} &=& -{\Ga_+}^a_{\;\;bc}\la^b\chi^c\no
\label{eq:Q_LR}
\end{eqnarray}
Local observables of the topological theory must take the following form
$$\cO_f \;=\; f_{a_1\cdots a_p;b_1\cdots 
b_q}\chi^{a_1}\cdots\chi^{a_p}\la^{b_1}\cdots\la^{b_q}$$
where $f$ is totally anti-symmetric in $a$'s as well as in $b$'s. Recall 
that $\chi\in\Ga(T_+^{0,1}), \la\in\Ga(T_-^{0,1})$, so one can regard $f$ 
as a section of $\Om_+^{0,p}(M)\otimes\Om_-^{0,q}(M)$. Here the subscripts 
$\pm$ remind us with respect to which complex structure the differential 
forms are decomposed.
Next we must require that $\cO_f$ be annihilated by the BRST operator 
$Q_L+Q_R$.
To write down the action of $Q_L$ on $\cO_f$, it is convenient to regard 
$f$ as a $(0,p)$ form for the complex structure $I_+$, with values in 
$\Om_-^{0,q}(M)$. A straightforward calculation gives
$$\{Q_L,\cO_f\} \;=\; \cO_{\bar{D}_{(+)}f}.$$
Here $\bar{D}_{(+)}$ is a covariantization of the ordinary Dolbeault 
operator $\bar{\part}$ corresponding to $I_+$. The covariantization uses 
the connection on $\Om_-^{0,q}(M)$ coming from the connection $\Ga_-$ on
$\TM$. On the other hand, one can regard $f$ as a $(0,q)$ form for $I_-$, 
taking values in $\Om_+^{0,p}(M)$. One gets
$$\{Q_R,\cO_f\} \;=\; \cO_{\bar{D}_{(-)}f},$$
where $\bar{D}_{(-)}$ now stands for a covariantization of Dolbeault 
operator $\bar{\part}$ for $I_-$
using the connection $\Ga_+$ on $\TM$.

The space of local observables has a natural bigrading by the left and 
right moving R-charges.
With respect to it, $Q_L$ has grade $(1,0)$, and $Q_R$ has grade $(0,1)$. 
The local observables fit into the following bicomplex:
\begin{equation*}
\begin{xy}
\xymatrix@C=10mm{
  & \vdots&\vdots&\vdots&\\
  \cdots\ar[r] &  \cO^{p-1,q+1}\ar[r]\ar[u]&\cO^{p,q+1}\ar[r]
            \ar[u]^{Q_R}&\cO^{p+1,q+1}\ar[r]\ar[u]&\ldots\\
  \cdots\ar[r]^{Q_L} &  \cO^{p-1,q}\ar[r]^{Q_L}\ar[u]&\cO^{p,q}\ar[r]^{Q_L}
            \ar[u]^{Q_R}&\cO^{p+1,q}\ar[r]^{Q_L}\ar[u]&\ldots\\
  \cdots\ar[r] &  \cO^{p-1,q-1}\ar[r]\ar[u]&\cO^{p,q-1}\ar[r]
            \ar[u]^{Q_R}&\cO^{p+1,q-1}\ar[r]\ar[u]&\ldots\\
  	&  \vdots\ar[u]&\vdots\ar[u]^{Q_R}&\vdots\ar[u]&\\
}
\end{xy}
\end{equation*}
The total cohomology of this bicomplex is the space of ``physical'' 
observables in our topological
theory. As usual, this means that there are two spectral sequences which 
converge to the BRST cohomology $H_{Q_{\rm BRST}}^*$. In practice, the 
computation is usually quite involved, unless the spectral sequences 
degenerate at a very early stage.

\subsection{Relation with twisted generalized complex structures}
In the special case $H=0$, it has been argued in \cite{Kap} that the BRST 
complex coincides with
the cohomology of the Lie algebroid $E_1$ associated with the generalized 
complex structure
$\J_1$. We will show that the statement is true for arbitrary $H$.

First let us recall the necessary definitions. A Lie algebroid, by 
definition, is a real vector bundle $E$ over
a manifold $M$ equipped with two structures: a Lie bracket $[\cdot,\cdot]$ 
on the space of smooth sections
of $E$, and a bundle morphism $a:E\ra\TM$, called the anchor map. These 
data satisfy two compatibility
conditions:
\begin{itemize}
\item[(i)] $a([s_1,s_2])=[a(s_1),a(s_2)]\; \forall s_1,s_2\in \Gamma(E)$, 
i.e. $a$ is a homomorphism of Lie
algebras.
\item[(ii)] $[f\cdot s_1,s_2]=f\cdot[s_1,s_2]-a(s_2)(f)\cdot s_1,\quad 
\forall
f\in C^\infty(M), \forall s_1,s_2\in\Gamma(E).$
\end{itemize}
If we take $E=\TM$, $a=id,$ and the bracket to be the ordinary commutator of 
vector fields, then both conditions are obviously satisfied. Thus a Lie
algebroid over $M$ should be thought of as a ``generalized tangent bundle''.
A complex Lie algebroid is defined similarly, except that $E$ is a complex vector
bundle, and $\TM$ is replaced with its complexification $\TM_\CC$.

There is an alternative, and perhaps more intuitive, way to think about 
Lie algebroids. For any vector bundle
$E$ we may consider a graded supermanifold $\Pi E$, i.e. the total space 
of the bundle $E$
with the fiber directions regarded as odd and having degree 1. It turns 
out that there is a one-to-one
correspondence between Lie-algebroid structures on $E$ and degree 1 odd 
vector fields $Q$ on $\Pi E$
satisfying $\{Q,Q\}=2 Q^2=0$~\cite{Vaintrob}. The correspondence goes as 
follows. Let $(x^b,\xi^\mu)$
be local coordinates on $\Pi E$, where $x^b$ are local coordinates on $M$, 
and $\xi^\mu$ are linear coordinates on the fiber. Any degree 1 odd vector 
field on $\Pi E$ has the form
$$
Q=a^b_\mu\xi^\mu\frac{\partial}{\partial 
x^b}+c^\mu_{\nu\rho}\xi^\nu\xi^\rho\frac{\partial}{\partial\xi^\mu},
$$
where $a^b_\mu$ and $c^\mu_{\nu\rho}$ are locally-defined functions on 
$M$. Let $e_\mu$ be the basis
of sections of $E$ dual to the coordinates $\xi^\mu$.
Define a map $a:E\ra \TM$ by
$$
a(e_\mu)=a^b_\mu \frac{\partial}{\partial x^b}
$$
and a bracket by
$$
[e_\nu,e_\rho]=c^\mu_{\nu\rho} e_\mu.
$$
One can show that these data define on $E$ the structure of a Lie 
algebroid if and only if $Q^2=0$.

Identifying functions on $\Pi E$ with sections of the graded bundle 
$\Lambda^\bul E^*$, we may regard $Q$
as a differential on the space of sections of this bundle. We will call the resulting complex 
the canonical complex of the Lie
algebroid, and its cohomology will be called the Lie algebroid cohomology.
Let us take, for example, $E=\TM$, with $a$ the 
identity map, and the standard Lie
bracket. Then the canonical complex is the complex of differential 
forms on $M$, with $Q$ being the usual de Rham differential, and the Lie 
algebroid cohomology is simply the de Rham cohomology of $M$.

To every twisted generalized complex manifold $(M,H,\J)$ one can associate 
a complex Lie algebroid by letting $E$ be the eigenbundle of $\J$ with 
eigenvalue $-\ii$. The bracket on $E$ is induced by the Courant
bracket on $\TM\op \TM^*$, and the anchor map is the projection to 
$\TM_\CC$. The associated complex
controls the deformations of the twisted generalized complex structure on 
$M$ (with $H$ fixed).
We claim that the BRST complex
discussed above is isomorphic to the Lie algebroid complex associated to 
the twisted generalized complex
manifold $(M,H,\J=\J_1)$.

To see the relation between the two complexes, let us define new fermionic 
coordinates:
\begin{equation}\label{newfermions}
\psi^a=\frac{1}{\sqrt2}(\psi^a_++i\psi^a_-),\quad 
\rho_a=\frac{1}{\sqrt2}g_{ab}(\psi^b_+-i\psi^b_-).
\end{equation}
One may regard $\psi$ and $\rho$ as fermion fields taking values in the 
pull-back of $\TM_\CC$ and $\TM^*_\CC$,
respectively. Their anti-commutation relations are
\begin{equation}\label{anticomm}
\{\psi^a,\psi^b\}=\{\rho_a,\rho_b\}=0,\quad 
\{\psi^a(\sigma),\rho_b(\sigma')\}=
\delta^a_b\delta(\sigma-\sigma').
\end{equation}
It is convenient to introduce a fermion field
$$
\Psi=\begin{pmatrix} \psi\\ \rho\end{pmatrix}
$$
taking values in $\TM_\CC\op\TM_\CC^*$.
In terms of $\Psi$, the anticommutation relations read
$$
\{\Psi(\sigma)^\alpha,\Psi(\sigma')^\beta\}=\left(q^{-1}\right)^{\alpha\beta}\delta(\sigma-\sigma'),
$$
where $q$ is the canonical scalar product on $\TM_\CC\op \TM_\CC^*$ which 
plays such a fundamental role
in generalized complex geometry.

It is easy to check that
$$
\begin{pmatrix} \chi+\lambda \\ g(\chi-\lambda)\end{pmatrix} 
=(1+\ii\J_1)\Psi
$$
Therefore any function of the bosonic coordinates $\phi^i$ and fermionic 
scalars $\chi,\lambda$ can be rewritten as
a function on $\Pi E$, where $E$ is the eigenbundle of $\J_1$ with 
eigenvalue $-\ii$. Thus the graded
vector spaces underlying the two complexes are naturally isomorphic. It 
remains to show that
the BRST differential $Q_{\rm BRST}$ coincides with the Lie algebroid differential $Q$. 
One could check this by
a direct computation,
but there is a more efficient route to this goal, with the added advantage 
of making the isomorphism
of the two complexes more obvious. To explain this indirect proof, we must 
first discuss
ground states in the Ramond-Ramond sector.

\section{Ramond-Ramond Ground States}

\subsection{Cohomology of states and differential forms}

So far we have been discussing the BRST cohomology of operators in the 
twisted theory. One may also
consider the BRST cohomology of states. In a topological field theory, 
there is a state-operator
isomorphism, so the two cohomologies are identical. In the physical 
(untwisted) SCFT, the cohomology
of operators is reinterpreted as the chiral ring, while the cohomology of 
states is reinterpreted
as the space of zero-energy states in the Ramond-Ramond sector. The 
isomorphism between these two
spaces is given by the spectral flow.

In this section we compute the space of ground states in the RR sector 
from scratch. There are several
reasons to do this. First, it may be interesting to consider $N=2$ 
sigma-models with H-flux
when the $U(1)_A$ charge is anomalous, i.e. the condition 
Eq.~(\ref{anomaly}) is not fulfilled, or more generally, when the twisted
generalized Calabi-Yau condition is not fulfilled.
Such theories cannot be topologically twisted, but both the chiral ring 
and the space of RR ground states
are perfectly well-defined and in general non-isomorphic. In the K\"ahler 
case ($H=0$), this is a familiar
situation: the chiral ring is given by $H^\bul(\Lambda^\bul TX)$, while 
the space of RR
ground states is $H^\bul(\Omega^\bul_X)$. Only in the Calabi-Yau case are 
the two spaces naturally
isomorphic. Second, if we use the point-particle approximation and replace 
the 2d sigma-model with
supersymmetric quantum mechanics (this approximation can be shown to be 
exact as far as RR ground states
are concerned), then the Hilbert space of the theory can be naturally 
identified with the
space of differential forms on $X$ (of all degrees). The supercharge 
becomes a differential operator
on forms, and can be easily computed. We will see that this operator is 
exactly the differential
associated to the twisted generalized complex structure $\J_1$ in 
Ref.~\cite{Gua}. {}From this one can infer without any computations that the 
chiral ring coincides with the Lie algebroid cohomology associated to 
$\J_1$. This is the result claimed in the end of the previous section,
except that we do not need to assume the existence of the topological 
twist.

Let us start by writing down the Noether charges associated with $Q_+$ and 
$Q_-$ in the point-particle
  approximation\footnote{We use the same symbols for the generators and 
their associated charges.}:
\begin{eqnarray}
\label{eq:Q_pm}
Q_+ &=& \psi_+^a g_{ab}\dot\phi^b - \frac{i}{6} 
H_{abc}\psi_+^a\psi_+^b\psi_+^c\no\\
Q_- &=& \psi_-^a g_{ab}\dot\phi^b + \frac{i}{6} 
H_{abc}\psi_-^a\psi_-^b\psi_-^c
\end{eqnarray}
Let $Q=Q_+ + iQ_-$ and $Q^*=Q_+-iQ_-$. The supersymmetry algebra implies 
that
$$Q^2={Q^*}^2=0, \qquad \{Q,Q^*\} = 4\cH$$
where $\cH$ is the Hamiltonian of the supersymmetric quantum mechanics:
$$
\cH=\frac12 g_{ab}\dot\phi^a \dot\phi^b -\frac14 R^{(+)}_{abcd}\psi_+^a\psi_+^b\psi_-^c\psi_-^d
$$
By the standard Hodge-de Rham argument, the supersymmetric ground states 
are in one-to-one correspondence with the elements of $Q$-cohomology.

The charge $Q$ can be thought of as an operator on differential forms via 
canonical quantization.
The classical phase space of the supersymmetric quantum mechanical system 
is $\TM\oplus \Pi(\TM\!\oplus\TM)$, where $\Pi(\TM\!\oplus\TM)$ is the 
parity reversal of $\TM\!\oplus\TM$. The two ``fermionic'' copies of $\TM$ 
come from $\psi_+$ and $\psi_-$. The symplectic form on $\TM$ is the 
standard one, while the symplectic form in the odd directions (which is 
actually symmetric) is given by the Riemannian metric $g$:
$$
\{\psi^a_\pm,\psi^b_\pm\}_{P.B.}=-ig^{ab},\quad 
\{\psi^a_\pm,\psi^b_\mp\}_{P.B.}=0.
$$
Canonical quantization identifies the Hilbert space with $L^2(S)$, the 
space of square-integrable sections of the spin bundle 
$S=S(\TM\!\oplus\TM)$. In the case at hand,
$\TM\oplus\TM$ has a natural complex polarization, using which the spin 
bundle $S$ can be identified with
$\wedge^\bullet(\TM^*)$. In other words, instead of $\psi_\pm$ we use the 
coordinates $\psi$ and $\rho$,
which can be quantized by letting $\psi^a$ be a wedge product with $dx^a$, 
and letting $\rho_b$ be a contraction with the vector field 
$\frac{\partial}{\partial x^b}$.

Now we discuss how $N=1$ supercharges $Q$ and $Q^*$ act on the Hilbert 
space.
Let us first consider the case $H=0$. Following the standard quantization 
procedure, one can easily
show that $Q=-i{\sqrt 2}\psi^a\nabla_a$, with $\nabla$ being the covariant 
derivative on the sections of the spin bundle $S(\TM\oplus\TM)$ that is 
induced from the Levi-Civita connection on $\TM$, and with $\psi^a$ acting 
as a Clifford multiplication. Under the isomorphism 
$S(\TM\!\oplus\TM)\simeq \wedge^\bullet(\TM^*)$, $Q$ is identified with 
the de Rham differential $d$, up to a factor $-i\sqrt 2$. 
This is the familiar statement that the 
space of
ground states in an $N=1$ supersymmetric quantum mechanics is isomorphic 
to the de Rham cohomology of
the target space. Now let us consider the case $H\neq 0$. Using 
(\ref{eq:Q_pm}) and (\ref{newfermions}) one can show that
\begin{eqnarray}
Q&=& -\sqrt2 i\psi^a\nabla_a +\frac{\sqrt2 
i}{6}H_{abc}\psi^a\psi^b\psi^c\no\\
Q^*&=& -\sqrt2 ig^{ab}\rho_b\nabla_a +\frac{\sqrt2 
i}{6}H^{abc}\rho_a\rho_b\rho_c\no
\end{eqnarray}
Up to a numerical factor $-\sqrt2 i$, $Q$ is identified with a twisted de 
Rham operator
$$d_H=d-H\wedge$$
while $Q^*$ is identified with its adjoint. Therefore the supersymmetric 
ground states are in one-to-one correspondence with the $d_H$-cohomology. 
This statement is also well-known~\cite{WR}.

It remains to identify the BRST operator, $\Qb$, in this context. The 
$R$-current is given by
$$J=-\frac{i}2\Big(\om_+(\psi_+,\psi_+)+\om_-(\psi_-,\psi_-)\Big),$$
under which $(1+iI_+)\psi_+$ and $(1+iI_-)\psi_-$ have charge $+1$ by 
canonical anticommutation relations. For our purpose, it will be more 
convenient to express $J$ in terms of the fermions $\psi,\rho$:
$$J=-\frac{i}2\big(\delta\om(\psi,\psi)-\al(\rho,\rho)-2\langle 
\tI\psi,\rho\rangle\big).$$
As discussed above, quantization amounts to substitutions:
$$\psi^a\leftrightarrow dx^a\wedge, \qquad \rho_a\leftrightarrow 
\i_{\part/\part x^a}\equiv \i_a.$$
Then the R-current is identified with the following operator on 
differential forms:
$$J\;=\; -i\big(\delta\om\wedge-\i_\al-\i_{\tI}\big)$$
where $\i_\al$ is the contraction with the Poisson bivector $\al$, and 
$\i_\tI$ is defined in a
local coordinate basis as
$$\i_\tI \;=\; \tI^a_{\;b}(dx^b\wedge)\circ \i_a.$$
Note that $\delta\om,\al,$ and $\tI$ can be read off $\J_1$ (see Eq.~(\ref{eq:J1J2})), and
therefore the operator $J$ depends only on the TGC-structure $\J_1$.

The BRST operator is given by
$$\Qb = \frac12\big(Q+[J,Q]\big).$$
Since $Q=d_H$, it is clear that $\Qb$ depends only on the 3-form 
$H$ and the twisted
generalized complex structure $\J_1$.
In the following two subsections, we will relate $\Qb$ to a differential operator on 
forms defined in Ref.~\cite{Gua}
as well as to the canonical complex of the Lie algebroid $E$.

\subsection{Differential forms on a twisted generalized complex manifold}

To proceed further, we need to discuss some properties of differential forms on a twisted generalized
complex manifold. Recall that on an ordinary complex manifold, the space of differential forms
is graded by a pair of integers, the first integer being the ordinary degree of a form, and the second 
integer being the difference between the number of holomorphic and antiholomorphic indices. If we think
of a form as a function on a supermanifold $\Pi \TM$, then the first integer is the eigenvalue of a
differential operator
$$
\deg=\theta^a\frac{\partial}{\partial\theta^a},
$$
while the second integer is the eigenvalue of
$$
-i\cdot\iota_I=-iI^a_{\;b}\theta^b\frac{\partial}{\partial\theta^a},
$$
where $I^a_{\;b}$ is the complex structure tensor. The existence of the second grading allows us to decompose the
de Rham differential $d$ into a holomorphic exterior differential $\partial$ and its antiholomorphic twin
$\bpartial$. Now, following Gualtieri~\cite{Gua}, we will define analogues of $-i\cdot\iota_I,\partial,$ and $\bpartial$ for twisted generalized complex manifolds.

Recall that $\TM\op \TM^*$ acts on $\Omega^\bul(M)$ via the spinor representation. Consider the subbundle
$E$ of $\TM_\CC\op \TM^*_\CC$ defined as the eigenbundle of the TGC-structure $\J$ with eigenvalue $-i$,
and its complex-conjugate $\bar E$. Since $E$ is isotropic with respect to the form $q$, one may regard
elements of $E$ as annihilation operators, and elements of $\bar E$ as creation operators, which
act on the fermionic Fock space $\Omega^\bul(M)$. In other words, the complex structure $\J$ on the
vector bundle $\TM\op \TM^*$ allows one to identify the Clifford algebra generated by $\TM_\CC\op \TM^*_\CC$
with the fermionic creation-annihilation algebra generated by $E\op \bar E$. In each fiber of $\Omega^\bul(M)$
we thus have a vacuum vector, defined up to a factor by the condition that $E$ annihilates it.
These vacuum vectors fit into a complex line bundle $U_0$ over $M$, which is obviously a subbundle
of $\Omega^\bul(M)$. We will call it the canonical line bundle of the TGC-manifold $(M,\J)$.
If $\J$ arises from an ordinary complex structure on $M$, then $U_0$ is the bundle of top-degree holomorphic
forms on the corresponding complex manifold. In general, $U_0$ does not have a definite degree (i.e. its
sections are inhomogeneous forms).

More generally, we can decompose $\Omega^\bul(M)$ into a direct sum
$$U_0\oplus U_1\oplus\cdots\oplus U_{2n}$$
where $U_0$ is the canonical line bundle for the TGC 
structure $\J$, and $U_k=\wedge^k {\bar E}\cdot U_0$. Fiberwise, this is simply a decomposition
of the fermionic Fock space into subspaces with a definite fermion number. Thus we obtained a grading
of the space of differential forms $\Omega^\bul(M)$ by a nonnegative integer $k$. In the case when $\J$ arises from an ordinary
complex structure $I$, this grading reduces to $i\cdot\iota_I + n$, where $n$ is the complex dimension of $M$.

The analogue of the de Rham operator $d$ is the ``twisted'' de Rham operator operator $d_H=d-H$.
The analogues of the Dolbeault operators $\partial$ and $\bpartial$ are defined as follows:
\begin{eqnarray}
\bar\part_H &=& \pi_{k+1}\circ d_H: \; \Ga(U_k)\to 
\Ga(U_{k+1})\no\\
\part_H &=& \pi_{k-1}\circ d_H: \; \Ga(U_k)\to \Ga(U_{k-1})\no
\end{eqnarray}

Note that all these constructions make sense even when $\J$ fails to be integrable with
respect to the twisted Courant bracket. It turns out that $\J$ is integrable with respect 
to the twisted Courant bracket if and only if
$d_H=\part_H+\bar\part_H$. This was proved in Ref.~\cite{Gua} in the case 
$H=0$, but one can easily
modify the argument so that it applies in general. For the sake of 
completeness, we provide a proof in the Appendix. Similarly, the Dolbeault operators
can be defined for almost complex manifolds, but the identity $d=\partial +\bpartial$ holds
if and only if the almost complex structure is integrable.

The twisted generalized Calabi-Yau condition that we mentioned above can be formulated in terms of the canonical line bundle $U_0$~\cite{Hitchin,Gua}. 
Namely, a TGC-manifold is called a TG-Calabi-Yau manifold if there exists a nowhere vanishing section $\Omega$ of $U_0$ which is $d_H$-closed. In view of the definition of $U_0$, this is the same as requiring
$$
\bpart_H\Omega=0.
$$
The topological part of the TG-Calabi-Yau condition (i.e. topological triviality of the line bundle $U_0$) 
is equivalent to $c_1(E)=0$. We have seen above that this condition ensures that the R-current necessary for twisting is nonanomalous. The second part (the existence of a ``twisted-holomorphic'' section $\Omega$)
is also quite important from the physical viewpoint: we will see that it ensures the absence of BRST-anomaly.

It is interesting to ask if the second part of the TG-Calabi-Yau condition follows from the first one.
The answer turns out to be negative, and this can be seen already for ordinary Calabi-Yau 
manifolds.\footnote{We are grateful to Misha Verbitsky for explaining this to us.}
Namely, if $c_1(M)=0$ and $M$ is not simply connected, it may happen that the canonical line bundle
is not trivial as a holomorphic line bundle (even though it is trivial topologically). In this case
there are no nowhere-vanishing holomorphic sections of the canonical line bundle.

\subsection{Cohomology of states and twisted generalized complex 
structures}
\label{sec:coh}

We are going to show that the BRST-cohomology of states is isomorphic
to the cohomology of $\bpart_H$ on differential forms on the TGC-manifold $(M,\J_1)$.
As a preliminary step, let us obtain a convenient explicit formula for the grading 
operator on $\Omega^\bul(M)$, defined in the previous subsection,
in terms of the twisted generalized complex structure $\J_1$.
Let $A=X+\xi\in \Gamma(\TM_\CC\oplus \TM^*_\CC)$. It can be regarded as an 
endomorphism of $\wedge^\bullet \TM_\CC^*$:
\begin{equation}\label{eq:act}
A\cdot \rho = \i_X\rho+\xi\wedge\rho.
\end{equation}
On the other hand, $\J_1$ is an endomorphism of
$\TM_\CC\oplus \TM_\CC^*$ with eigenvalues $\ii$ and $-\ii$. By definition, the 
grading operator $R(\J_1)$
must satisfy
$$
[R(\J_1),A]=-i \J_1 A,\quad \forall A\in \Gamma(\TM_\CC\oplus \TM_\CC^*).
$$
Obviously, this condition determines $R(\J_1)$ up to a constant.
Using the explicit matrix form (\ref{eq:J1J2}) of $\J_1$ , one gets
$$\J_1A\cdot\rho = \i_{\tI X} + \i_{\al(\xi)} - \i_X\delta\om\wedge  - 
\tI^t(\xi)\wedge.$$
Then it is easy to check that the following is a solution to the above 
equation:
$$R(\J_1) = -\ii(\delta\om\wedge - \i_\al-\i_\tI).$$
The general solution may differ from this only by a constant.

{}From the result of the last section, one immediately sees that under the 
identification of $Q\leftrightarrow d_H$ and the identification of the 
Hilbert space as the space of differential forms, the R-current generator 
$J$ is identified as
$$J\;\leftrightarrow R(\J_1)+{\rm const.}$$
The BRST operator $\Qb$ then becomes
\begin{eqnarray}
\Qb &=& \frac12\big(Q+[J,Q]\big)\no\\
 	&=& \frac12\big(d_H +[R(\J_1),d_H]\big)\no\\
 	&=& \bar\part_H\no
\end{eqnarray}
This is the desired result.

Now we can show that the BRST-cohomology of operators is isomorphic to the Lie
algebroid cohomology of $E$.
Recall that the $\bpart_H$ complex is a differential graded module 
over the Lie algebroid complex
$(\Lambda^\bul {\bar E},d_L)$, i.e. the following identity holds for any 
section $s$ of $\Lambda^\bul {\bar E}$
and any differential form $\rho$:
$$
\bpart_H (s\cdot\rho)=(d_L s)\cdot\rho+(-1)^{|s|}s\cdot \bpart_H\rho.
$$
This identity is proved in Ref.~\cite{Gua} for the case $H=0$, but the proof is 
valid more generally.
Since we identified the space of sections of $(\Lambda^\bul {\bar E},d_L)$ with 
the space of
operators, the space of forms with the Hilbert space of the SUSY quantum 
mechanics, and
$\bpart_H$ with the representation of the BRST charge on the Hilbert 
space, it follows
that
$$
[\Qb,s]=d_L s.
$$
This implies that the cohomology of $d_L$ is isomorphic to the BRST cohomology of operators,
as claimed.

\section{Topological correlators and the Frobenius structure}
For any $N=2$ $d=2$ field theory we may consider the chiral ring, as well 
as the cohomology
of the supercharge $Q_L+Q_R$ on the states in the Ramond sector. The 
latter is a module over
the former. The two spaces are not isomorphic in general. But if the 
theory admits a topological
B-twist, the two spaces are always isomorphic, by virtue of the 
state-operator correspondence
in a topological field theory. More precisely, the space of states of a 2d 
TFT is an algebra
with  a nondegenerate scalar product $(\cdot,\cdot)$ such that
$$
(a,bc)=(ab,c).
$$
Such algebras are called Frobenius. All topological correlators can be 
expressed in terms of the
Frobenius structure on the space of states. For example, genus zero 
correlators are given by
$$
\langle a_1\ldots a_n\rangle_{g=0}=(a_1,a_2\ldots a_n).
$$

Consider now an $N=2$ sigma-model for which the condition 
Eq.~(\ref{anomaly}) is satisfied, and the $U(1)_A$ R-charge is nonanomalous.
One expects that the theory admits a topological B-twist, and therefore the
chiral ring, which is known to be isomorphic to the Lie algebroid cohomology of $E_1$,
is a supercommutative Frobenius algebra. In fact, we will see that in order for
a BRST-invariant measure in the path-integral to exist, the target manifold must be
a TG-Calabi-Yau manifold, which is stronger than Eq.~(\ref{anomaly}).

Note that the Frobenius scalar product $(\cdot,\cdot)$
can be recovered from the ``trace'' function:
$$
\Tr(a)=(1,a)
$$
by letting $(a,b)=\Tr(ab).$ The name ``trace'' is used because $\Tr$ 
vanishes on commutators
(in the graded case, on graded commutators). Let $\Omega$ be a 
$\bpart_H$-closed differential form
which sits in the component $U_0$. For a twisted generalized Calabi-Yau 
such a form exists and
is unique up to a constant factor. Note that $\Omega$ is also 
$d_H$-closed. Consider now a bundle automorphism $p:\TM\op \TM^*$ which looks as follows:
$$
p:(v,\xi)\mapsto (v,-\xi),\quad \forall v\in\Gamma(\TM), \forall \xi\in \Gamma(\TM^*).
$$
This automorphism takes the form $q$ to $-q$, and maps the Courant bracket twisted by $H$ to
the Courant bracket twisted by $-H$. It follows from this that for any twisted generalized
complex structure $\J$ the bundle map $\J'=p^{-1}\J p$ is also a twisted generalized complex
structure, for the opposite H-field $H'=-H$. Moreover, it is easy to see that $(M,-H,\J')$ is a twisted
generalized Calabi-Yau if and only if $(M,H,\J)$ is one. (From the physical viewpoint, $p$ corresponds
to worldsheet parity transformation, and the above facts are obvious). In particular, we have
a decomposition
$$
\wedge^\bullet \TM^*\otimes\CC = U'_0\oplus U'_1\oplus\cdots\oplus 
U'_{2n}
$$
Let $\Omega'$ be the
$\bpart_{H'}$-closed differential form which sits in the component $U'_0$. We claim that the
trace function on the Lie algebroid cohomology is given by
$$
\Tr(\alpha)\sim\int_M \Omega'\wedge \alpha\cdot\Omega,
$$
where $\alpha$ is a $d_L$-closed section of $\Lambda^\bul(E_1^*)$. 

To derive this formula, we recall that
the Frobenius trace is computed by the path-integral on a Riemann sphere with an insertion of the
operator corresponding to $\alpha$. Since we are dealing with a topological theory, we must also turn
on a $U(1)$ gauge field coupled to the R-current participating in the twisting. This gauge field must
be equal to the spin connection, which means that the total flux through the sphere is $2\pi$.
Let us stretch the sphere into a long and thin cigar, so that the insertion point of $\alpha$ is 
somewhere in the
middle portion. The value of the path-integral does not change, of course, but it may now
be evaluated more easily by reducing the computation to the supersymmetric quantum  mechanics.
The path-integral on each hemisphere gives a state in the Ramond-Ramond sector, which may
be approximated in the point-particle limit by a function of the zero-modes. Bosonic zero-modes
are simply coordinates on $M$, while fermionic zero-modes are $\psi^i$, taking values
in $\TM_\CC$. Thus the Ramond-Ramond state is represented by a function on $\Pi \TM_\CC$, i.e.
by a (complex-valued) differential form. We have described above how $\alpha$ acts on differential
forms. 

It remains to identify the particular RR states arising from performing the path-integral
over each hemisphere, and then integrate over the zero-modes. Since we are not inserting any 
operators on the hemispheres, the RR ground state in question is the 
spectral flow of the unique vacuum state in the NS sector, and therefore in the point-particle
approximation is represented by the form $\Omega$ defined above.\footnote{If a $d_H$-closed section
of the line bundle $U_0$ does not exist, then supersymmetry is spontaneously broken, i.e. there are no
RR states with zero energy. {}From the point of view of the topological theory, this means that the measure
in the path-integral fails to be BRST invariant, i.e. there is a BRST anomaly.}
However, there is a subtlety related to the choice of orientation. This subtlety arises because
our identification of RR states with differential forms depends on orientation: exchanging left-moving
and right-moving fermions is equivalent to performing a Hodge duality on forms. In the physical
language, Hodge duality is simply the Fourier transform of fermionic zero-modes. If we induce the
orientations of both hemispheres from a global orientation of the Riemann sphere, 
then the wave-function coming from one 
hemisphere will be given by a function of the fermionic zero-modes $\psi^i$, while the wave-function 
from the other hemisphere will be a function of the Fourier-dual zero-modes. In order to evaluate the 
path-integral one first has to Fourier transform the second state, and only then multiply the wave-functions 
and integrate over the zero-modes. Alternatively, we can choose the opposite orientation for the second
hemisphere, so that there is no need for Fourier transform. This also requires flipping the
sign of $H$, since the worldsheet theory is not parity-invariant. We conclude that the wave-function
from the second hemisphere is given by $\Omega'$, and the path-integral in question
is given by
$$
\Tr(\alpha)\sim\int_M \Omega'\wedge \alpha\cdot\Omega.
$$

Let us check that this formula is BRST-invariant, i.e. that 
it vanishes if $\alpha=d_L\beta$ for some $\beta$. Indeed, we have
$$
\Tr(d_L\beta)=\int_M \Omega'\wedge\bpart_H (\beta\cdot\Omega)=\int_M 
\Omega'\wedge (d_H+[R(\J_1),d_H])
(\beta\cdot\Omega).
$$
Next we have to use the following two identities valid for any two forms $\gamma,\eta$:
\begin{align}
\int_M \gamma\wedge d_H\eta &=-(-1)^{|\gamma|}\int_M (d_{H'}\gamma)\wedge\eta\\
\int_M \gamma\wedge R(\J_1)\eta &=-\int_M (R(\J'_1)\gamma)\wedge\eta,
\end{align}
where $H'=-H$, and $\J'=p^{-1}\J p$. Then we get
\begin{multline}
\Tr(d_L\beta)=-(-1)^{|\Omega'|}\int_M \left((d_{H'}+[R(\J'_1),d_{H'}])\Omega'\right)\wedge \beta\cdot\Omega\\
=-(-1)^{|\Omega'|}\int_M \bpart_{H'}\Omega'\wedge\beta\cdot\Omega=0.
\end{multline}

Let us also check that this formula reduces to the known expressions in the 
case of the ordinary
A and B-models with $H=0$ and $I_+=I_-$. For the ordinary B-model, $\J_1'=\J_1$, 
$\Omega'=\Omega$, and the form $\Omega$ is simply
the top holomorphic form on $M$. It is obvious that our formula for the trace function reduces
to the standard formula for the B-model~\cite{WittenMirr}.
For the A-model, the situation  is more interesting.
The relevant generalized complex structure is $\J_2$, and we have $\J_2'=-\J_2$. The forms $\Omega$
and $\Omega'$ are given by
$$
\Omega=e^{i\omega},\quad \Omega'=e^{-i\omega}.
$$
The complex Lie algebroid for the A-model is isomorphic to $\TM_\CC$, thus the Lie algebroid cohomology
is isomorphic to the complex de Rham cohomology. The usual formula for the Frobenius trace on 
$H^\bul(M)$ is
$$
\Tr(\beta)=\int_M\beta,\quad \beta\in \Omega^\bul(M), d\beta=0.
$$
This does not seem to agree with our formula. But one should keep in mind that the identification between
the Lie algebroid cohomology and de Rham cohomology is nontrivial, and as a result, although the bundle
$\Lambda^\bul E_2^*$ is isomorphic to $\Omega^\bul(M)$, the action of $\Omega^\bul(M)$ on itself
coming from the action of $\Lambda^\bul E_2^*$ on $\Omega^\bul(M)$ is not given by the wedge product.
To describe this action, let us identify the space of sections of
$\Omega^\bul(M)$ with the graded supermanifold $\Pi \TM$. Let $\alpha\in \Omega^k(M)$ be given by
$$
\alpha=\frac{1}{k!}\alpha_{a_1\ldots a_k} dx^{a_1}\wedge\ldots \wedge dx^{a_k}.
$$
The action we are after is obtained by associating to $\alpha$ the following differential operator on $\Pi \TM$:
$$
\frac{1}{k!}\alpha_{a_1\ldots a_k} D^{a_1}\ldots D^{a_k},
$$
where
$$
D^a=\theta^a+i\left(\omega^{-1}\right)^{ab}\frac{\partial}{\partial\theta^b}.
$$
The operators $D^a$ anticommute, so this is well-defined. On the other hand, in the usual description of
the A-model, the action of $\Omega^\bul(M)$ on itself is given by the ordinary wedge product (plus quantum
corrections, which we neglect in this paper).

The difference between our description of the A-model and the usual one is due to a different identification of 
the fermionic fields with operators on forms. While we identified $\psi^a$ with ``creation'' operators
$dx^a$ and $\rho_a$ with ``annihilation'' operators, the usual identification is different:
$$
\psi_+^\bi\mapsto dx^\bi,\quad \psi_-^i\mapsto dx^i,\quad g_{\bi j}\psi_+^j\mapsto \i_{\frac{\partial}{\partial x^\bi}},
\quad g_{j\bi}\psi_+^\bi\mapsto \i_{\frac{\partial}{\partial x^j}}.
$$
This choice is related to ours by a Bogolyubov transformation. In the usual description the vacuum state with the
lowest R-charge $J_L-J_R$ is given by the constant 0-form on $M$. It is easy to see that the Bogolyubov transformation 
maps it to the inhomogeneous form $e^{i\omega}$. The same transformation also maps the ordinary degree of
a differential form to the nonstandard grading on $\Omega^\bul(M)$ defined in Ref.~\cite{Gua} and explained above.
Thus our formula agrees with the standard one after a Bogolyubov transformation (and if one neglects quantum corrections).

\section{Towards the Twisted Generalized Quantum Cohomology Ring}

So far we have only discussed the classical ring structure on the space of topological observables. In general, the actual
ring structure is deformed by quantum effects. A well-known example is the ordinary A-model, whose ring of BRST invariant 
observables (the quantum cohomology ring) is a deformation of the de Rham cohomology ring $H^\bul(M,\CC)$ induced 
by worldsheet instantons. In this section, we carry out the analysis for generic twisted topological sigma model with $H$-flux,
and identify worldsheet instantons which can contribute to the deformation of the ring structure. This section is essentially 
an extension of the analysis of Section 8.2 of Ref.~\cite{Kap} to the case $H\neq0$.

As is well-known, the path integral of a (cohomological) TFT can be localized around the $Q_{\rm BRST}$ invariant field 
configurations \cite{WittenMirr}. For our generalized B-model, the BRST variations of some of the fields are already given 
in Eq.~(\ref{eq:Q_LR}). We will also need the following BRST transformations for $(1-iI_\pm)\psi_\pm$:
\begin{eqnarray}
\Big\{Q_{\rm BRST},\frac12(1-iI_+)\psi_+\Big\} &=& \frac{i}2(1-iI_+)\part_+\phi + \cdots\no\\
\Big\{Q_{\rm BRST},\frac12(1-iI_-)\psi_-\Big\} &=& -\frac{1}2(1-iI_+)\part_-\phi + \cdots\no
\end{eqnarray}
where the dots involve fermion bilinear terms. The BRST invariant configurations are given by setting the fermionic fields 
$\psi_\pm$ to zero and demanding 
$$\frac{1}2(1-iI_+)\part_+\phi = 0, \qquad \frac{1}2(1-iI_-)\part_-\phi = 0.$$
In terms of the generalized complex structure, the above equation is equivalent to
\begin{equation}
\frac12(1-i\J_1)\left(\begin{array}{c}\part_1\phi\\ g\part_0\phi\end{array}\right)=0.
\label{eq:instanton}
\end{equation}
This is the same instanton equation as obtained in Ref.~\cite{Kap} in the case of $H=0$, and the results of \cite{Kap} carry over to our 
case. For reader's convenience, we summarize them below.

To find Euclidean instantons, we Wick-rotate $\part_0\to \sqrt{-1}\part_2$ as in Ref.~\cite{Kap}. 
Eq.~(\ref{eq:instanton}) then leads
to the following equations
\begin{equation}\label{eq:genhol}
\begin{array}{lll}\tilde\om\partial_1\phi=0, &\qquad & \partial_1\phi =-\delta I\partial_2\phi\\ 
  \tilde\om\partial_2\phi  =0,& \qquad & \partial_2\phi =\delta I \partial_1\phi.\end{array}
\end{equation}
It is not difficult to see that the solutions to Eqs.~(\ref{eq:genhol}) are ``twisted generalized 
holomorphic maps'' with respect to $\J_2$. The precise meaning of this is as follows. 
The differential $d\phi$ composed with the natural embedding 
$j:\TM\to\TM\oplus\TM^*$ defines a map
$$j\circ d\phi : T\Sigma\to \TM\oplus\TM^*.$$
Eq.~(\ref{eq:genhol}) then says that $j\circ d\phi$ intertwines the TGC structure $\J_2$ and the complex structure 
on the worldsheet $I_\Sigma$. That is,
$$\J_2(j\circ d\phi) = (j\circ d\phi)I_\Sigma$$
Solutions of this equation generalize both the
holomorphic maps of the ordinary A-model and the constant maps of the ordinary B-model. 

On general grounds, the ring structure 
of topological observables may admit non-trivial quantum corrections coming from these worldsheet 
instantons. We will not try to describe these corrections more precisely here. But note that
for a generic TGC-structure $\J_2$ the TG-holomorphic instanton equation is much more restrictive than the 
ordinary holomorphic instanton equation. Indeed, it requires the image of $T\Sigma$ under $d\phi$ to lie
in the kernel of the map $\tilde\omega$. For a generic $\J_2$ and at a generic point of $M$ the 2-form
$\tilde\omega$ is nondegenerate, and so this condition does not allow nonconstant instantons. In other words,
all nontrivial instantons must be contained in the subvariety where $\tilde\omega$ is degenerate. The extreme cases are the ordinary B-model, where $\tilde\omega$ is a symplectic form and there are no nontrivial instantons, 
and the ordinary A-model, where $\tilde\omega$ vanishes identically.

\section{Discussion}

In this paper we have studied the topological sector of $(2,2)$ sigma-models with $H$-flux. We found that
the results are very conveniently formulated in terms of twisted generalized complex structures.
For example, the chiral ring is isomorphic (on the classical level) to the cohomology of a certain
Lie algebroid which controls the deformation theory of a twisted generalized complex structure.
On the quantum level, the two rings are isomorphic as vector spaces, but the ring structures may be
different due to worldsheet instantons. It would be interesting to further study these quantum corrections.
In particular, we expect that the quantum ring structure depends only on one of the two twisted generalized
complex structures present. (This is the analogue of the statement that the quantum cohomology ring is
independent of the choice of the complex structure~\cite{WittenMirr}.) To prove this, one has to show that varying the TGC-structure $\J_2$ changes the action of the sigma-model by BRST-exact terms.

It is expected on general grounds that the moduli space of $N=2$ SCFTs is a product of two spaces, 
corresponding to deformations by elements of the (c,c) and (a,c) rings. It follows from our work that
for $N=2$ sigma-models with $H$-flux these two moduli spaces are identified with the moduli spaces of
two independent twisted generalized complex structures $\J_1$ and $\J_2$. {}From the mathematical viewpoint,
this means that the deformation theory of twisted generalized Calabi-Yau manifolds is unobstructed.
It would be very interesting to prove this rigorously.

Recall that the well-known K\"ahler identities
$$
\partial \partial^*+\partial^*\partial=\bpartial \bpartial^*+\bpartial^*\bpartial=\frac12 (d d^*+ d^* d),
$$
are interpreted physically as $N=2$ supersymmetry relations. It follows from the results of this paper
that for twisted generalized K\"ahler manifolds analogous relations hold true:
$$
\partial_H \partial_H^*+\partial_H^*\partial_H=\bpartial_H \bpartial_H^*+\bpartial_H^*\bpartial_H=
\frac12 \left(d_H d_H^*+ d_H^* d_H\right).
$$

It should be clear that all the notions pertaining to ordinary $N=2$ sigma-models make sense when one
allows for the possibility of $H$-flux. For example, a pair of twisted generalized Calabi-Yau
manifolds $M$ and $M'$ are called mirror if the generalized A-model of $M$ is isomorphic to the
generalized B-model of $M'$, and vice versa. In mathematical terms, this means that the Frobenius
manifold corresponding to deformations of the twisted generalized complex structure $\J_1$ on $M$
is isomorphic to the Frobenius manifold corresponding to deformations of $\J_2$ on $M'$, and vice versa.
Further, it is possible to define the categories of generalized A and B-branes, and one expects
that mirror symmetry exchanges them. The geometry of generalized A and B-branes deserved further study;
initial steps in this direction have been made in Refs.~\cite{Kap,Zabzine}.

It would be very interesting to find examples of mirror pairs of twisted generalized Calabi-Yau
manifolds. There is a slight problem here though: we do not expect any {\it compact} examples of
twisted generalized Calabi-Yau manifolds with $H\neq 0$ to exist. If such an example existed, it
would give rise to a superconformal $N=2$ sigma-model with integral central charge. By rescaling the
metric and the $H$-field (so that the volume of the manifold is large), 
we would get a metric and an $H$-field
on $M$ which satisfy supergravity equations of motion. But there are well-known theorems that force
all such smooth supergravity solutions to have zero $H$-field~\cite{MN,IP,GKP}.

Note in this connection that the simplest nontrivial example of a TG K\"ahler manifold $M=S^3\times S^1$
(see Ref.~\cite{Gua}) is not a TG-Calabi-Yau manifold. Even though the topological 
condition $c_1(E)=0$ is trivially satisfied
($M$ has no cohomology in degree two), the line bundle $U_0$ does not have a section which is 
$d_H$-closed.\footnote{We are grateful to M. Gualtieri for explaining this to us.}

Thus to study mirror symmetry for twisted generalized K\"ahler manifolds we either have to work with
noncompact manifolds, or drop the Calabi-Yau condition. Both possibilities are interesting.
What we are lacking at present is a generalization of the K\"ahler quotient (or toric geometry)
construction. This would provide us with a large supply of TG K\"ahler manifolds and perhaps would also
enable us to find their mirrors (cf.~\cite{HV}). We plan to return to this subject in the future.

\section*{Appendix}
A twisted generalized almost complex structure is defined just like
a twisted generalized complex structure, except that the last condition
(integrability of $E=\ker (\J+i)$ with respect to the Dorfman bracket) is dropped.
Given a TG almost complex structure $\J$ on $(M,H)$, one can define operators $\part_H$
and $\bpart_H$ on differential forms on $M$ (see Section 4.2).
In this appendix we prove the following integrability criterion for $\J$:
\vspace{4mm}

\noindent{\bf Theorem.} The twisted generalized almost complex structure 
$\J$ is integrable if and only if $d_H=\part_H+\bar{\part}_H$.
\vspace{5mm}

We shall outline a proof of the theorem following 
Gualtieri, who proved it in the special case $H=0$ \cite{Gua}. Let $\rho$ 
be an arbitrary differential form, and let $A=X+\xi$, $B=Y+\eta$ be 
arbitrary sections of $E$. It is straightforward to show (using 
Eq.~(\ref{eq:act}) and the Cartan identities $\cL_X = \iota_X\circ 
d+d\circ\iota_X$, $\iota_{[X,Y]}=[\cL_X,\iota_Y]$) that
\begin{eqnarray}
\label{eq:AB1}
A\cdot B\cdot d\rho &=& d(BA\rho)+B\cdot d(A\rho)-A\cdot 
d(B\rho)+[A,B]\rho\\
A\cdot B\cdot(H\wedge\rho) &=& -\iota_Y\iota_X H\wedge\rho+\iota_Y 
H\wedge(A\rho)\no\\
 	&& -\iota_X H\wedge(B\rho)+H\wedge(AB\rho)
\label{eq:AB2}
\end{eqnarray}
Subtracting (\ref{eq:AB2}) from (\ref{eq:AB1}), one obtains
\begin{equation}
A\cdot B\cdot d_H\rho = d_H(BA\rho)+B\cdot d_H(A\rho)-A\cdot 
d_H(B\rho)+[A,B]_H\cdot\rho.
\label{eq:AB}
\end{equation}

The rest of the proof now follows exactly as in Ref.~\cite{Gua}. First let's 
assume $\J$ is integrable. For $\rho\in \Ga(U_0)$, (\ref{eq:AB}) 
reduces to $AB\cdot d_H\rho=[A,B]_H\cdot\rho=0$. Since $d_H\rho$ has no 
component in $U_0$, it follows that $d(\Ga(U_0))\subset 
\Ga(U_1)$ and thus $d_H=\part_H+\bar\part_H$ holds for $\rho\in 
\Ga(U_0)$. Now assume $d_H=\part_H+\bar\part_H$ holds for all $U_k, 
0\le k<i$, and let $\rho\in \Ga(U_i)$ and $A,B\in \Ga(E)$ as 
before. Equation (\ref{eq:AB}) now shows that $AB\cdot d_H\rho\in 
\Ga(U_{i-3}\oplus U_{i-1})$, which in turn implies $d_H\rho \in 
\Ga(U_{i-1}\oplus U_{i+1})$. By induction, one concludes that 
$d_H=\part_H+\bar\part_H$ on $\wedge^\bullet \TM^*\otimes\CC$. The 
converse is also true by similar argument.

\section*{Acknowledgments}
A.K. is grateful to Andrew Frey, Marco Gualtieri, and Misha Verbitsky for advice. 
Y.L. would like to thank Vadim Borokhov, Takuya Okuda, and Xinkai Wu for
helpful discussions. This work was supported in part by the DOE grant DE-FG03-92-ER40701.


\begin{thebibliography}{99}

\bibitem{BSVV}
J.~Bogaerts, A.~Sevrin, S.~van der Loo and S.~Van Gils,
``Properties of Semi-Chiral Superfields,''
Nucl.\ Phys.\ B {\bf 562} (1999) 277
[arXiv:hep-th/9905141].

\bibitem{GHR} S.~J.~Gates, C.~M.~Hull and M.~Ro\v{c}ek, ``Twisted Multipliets and 
New Supersymmetric Nonlinear Sigma Models,''Nucl. Phys. B {\bf 248} (1984) 157.

\bibitem{GKP}
S.~B.~Giddings, S.~Kachru and J.~Polchinski,
``Hierarchies from fluxes in string compactifications,''
Phys.\ Rev.\ D {\bf 66} (2002) 106006
[arXiv:hep-th/0105097].

\bibitem{GMPT}
M.~Grana, R.~Minasian, M.~Petrini and A.~Tomasiello,
``Supersymmetric backgrounds from generalized Calabi-Yau manifolds,''
JHEP {\bf 0408}, 046 (2004)
[arXiv:hep-th/0406137].

\bibitem{Gua} M.~Gualtieri, ``Generalized Complex Geometry,'' D.Phil 
thesis, Oxford University [arXiv:math.DG/0401221].

\bibitem{Hitchin} N.~Hitchin, ``Generalized Calabi-Yau Manifolds,'' 
Q.\ J.\ Math. {\bf 54} (2003) 281 [arXiv:math.DG/0209099].

\bibitem{HV}
K.~Hori and C.~Vafa, ``Mirror symmetry,'' arXiv:hep-th/0002222.


\bibitem{IKR}
I.~T.~Ivanov, B.~b.~Kim and M.~Ro\v{c}ek,
``Complex Structures, Duality and WZW Models in Extended Superspace,''
Phys.\ Lett.\ B {\bf 343} (1995) 133 [arXiv:hep-th/9406063].

\bibitem{IP}
S.~Ivanov and G.~Papadopoulos,
``A no-go theorem for string warped compactifications,''
Phys.\ Lett.\ B {\bf 497}  (2001) 309.
[arXiv:hep-th/0008232].

\bibitem{Kap} A.~Kapustin, ``Topological Strings on Noncommutative 
Manifolds,'' arXiv:hep-th/0310057.

\bibitem{Lind1} U.~Lindstrom,
``Generalized N = (2,2) supersymmetric non-linear sigma models,''
Phys.\ Lett.\ B {\bf 587}, 216 (2004)
[arXiv:hep-th/0401100].

\bibitem{Lind2} U.~Lindstrom, R.~Minasian, A.~Tomasiello and M.~Zabzine,
``Generalized complex manifolds and supersymmetry,''
arXiv:hep-th/0405085.


\bibitem{LZ}
S.~Lyakhovich and M.~Zabzine, ``Poisson Geometry of Sigma Models with 
Extended Supersymmetry,'' arXiv:hep-th/0210043.

\bibitem{MN}
J.~M.~Maldacena and C.~Nunez,
``Supergravity description of field theories on curved manifolds and a no go
theorem,'' Int.\ J.\ Mod.\ Phys.\ A {\bf 16}, 822 (2001)
[arXiv:hep-th/0007018].

\bibitem{Rocek} M.~Ro\v{c}ek,
``Modified Calabi-Yau Manifolds with Torsion,'' in {\em Essays on Mirror 
Manifolds}, pp. 480-488, ed. S.~T.~Yau, International Press, Hong Kong, 1992.

\bibitem{Royt} D. Roytenberg, 
``On the structure of graded symplectic supermanifolds and Courant algebroids,''
in {\em Quantization, Poisson brackets and beyond (Manchester, 2001)}, pp. 169-185, Contemp. Math., 
{\bf 315},  AMS (2002) [arXiv:math.SG/0203110].


\bibitem{WR} R.~Rohm and E.~Witten,
``The Antisymmetric Tensor Field In Superstring Theory,''
Annals of Phys.\  {\bf 170} (1986) 454.

\bibitem{Vaintrob} A.~Yu.~Vaintrob, ``Lie algebroids and homological 
vector fields,'' Uspekhi Mat. Nauk {\bf 52} (1997) 161-162.

\bibitem{WittenTop} E.~Witten, ``Topological Quantum Field Theories,'' 
Comm. Math. Phys. 117 (1988) 353.

\bibitem{WittenMirr} E.~Witten, ``Mirror Manifolds and Topological Field 
Theory,'' in {\em Essays on Mirror 
Manifolds}, pp. 120-158, ed. S.~T.~Yau, International Press, Hong Kong, 1992
[arXiv:hep-th/9112056].

\bibitem{Zabzine} M.~Zabzine, ``Geometry of D-branes for general N = (2,2) sigma models,''
arXiv:hep-th/0405240.









\end{thebibliography}
\end{document}